\documentclass[12pt]{article}
\usepackage[dvips]{epsfig}

\begin{document}
\thispagestyle{empty}
\vspace*{1cm}
\begin{flushright}
{\bf ISU-IAP.Th 98-07, Irkutsk}
\end{flushright}
\vspace*{1cm}

\begin{center}
{\Large  Compton scattering by pion: there is no room
for the off--shell effects.\\[1.5cm]
A.E.Kaloshin \footnote{EM:\ kaloshin@physdep.isu.runnet.ru}  } \\[1cm]
{\sl  Institute of Applied Physics,
Irkutsk State University,\\  blvd Gagarin 20, 664003 Irkutsk, Russia and \\
Department of Theoretical Physics,\
Irkutsk State University
}\\[2cm]
\end{center}
\vspace{0cm}

\begin{center}
{\Large \bf Abstract}
\end{center}
\vspace{1cm}

We show that the off--shell contributions in Compton scattering by
pion may exist only in the single exchange diagrams. These contributions
are canceled completely in the total gauge--invariant amplitude as
it confirmed  by  one--loop calculations. It explains, in particular,
some results of the chiral pertubation theory in the order  $p^4$
but this cancelation has no relation to chiral symmetry.
\newpage

\section{Introduction}

\indent

The problem of incorporating of the internal structure of
particles in electromagnetic (~electroweak~)
processes has different aspects and a long history.
Here we would like to discuss one particular
question from this area concerned  the  off--shell effects
in Compton scattering by a pion.
First of all to attach the exact meaning to above phrase
let's look  at the Born diagrams of Fig. 1.
\begin{figure}[h]
\begin{center}
\includegraphics*[height=3.5cm]{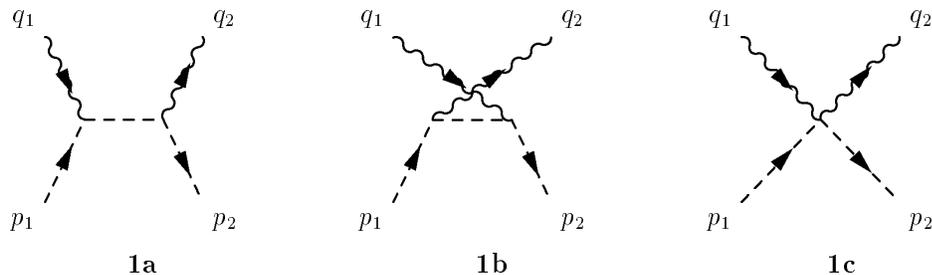}
\end{center}
\caption{Born contributions for Compton scattering}
\label{Born}
\end{figure}

The internal structure of pions
and strong interactions can  significantly modify
the amplitude, nevertheless
the threshold theorem of Low \cite{Low} guarantees
that the Born contribution dominates near the threshold.
Corrections to the Born terms (polarizabilities and the higher structure
constants in the threshold expansion) are defined by different
physical effects.
\footnote{ Some details and history of the low--energy theorem
including the structure corrections may be found in \cite{Pet}.}
The so called off--shell effects arise because
the exchange pion in diagrams of Fig. 1 is virtual.
It leads  to appearance in vertices of two form--factors
which depend on the virtuality and
can, generally speaking, modify the amplitude.

The subject under discussion is directly related
with two--photon experiments on  $e^+e^-$ colliders
studying
$\gamma\gamma\to\pi^+\pi^-$ reaction near threshold.
In the framework of S--matrix description for this process
(see, for instance, \cite{MP,KS}) the off--shell contributions
usually are not included (~or one should believe that they
are properly absorbed  by existing arbitrary parameters~).
But there exist some attempts
(e.g. \cite{CELLO})
of simple accounting (parametrization) of these effects.

From more theoretical point of view,
the contributions under discussion were considered
\cite{Sher}  in the framework of chiral perturbation theory.
It was found that there exists few effective lagrangians
(~physically equivalent, the authors call them representations~)
which generate the same Compton amplitudes
at different off--shell formfactors in the electromagnetic
vertex. The authors came to conclusion that
the off--shell effects are not only model dependent but also
representation dependent.

However the results of \cite{Sher}
(~recall that  they are dealing with the momentum
expansion up to the order~~$p^4$~)
lead to thought about total disappearance of
the off--shell contributions by some reasons.
It turns out that this is truth and the reasons are rather simple.

Our main point may be easily clarified without any formulae.
The off--shell effects in a vertex and propagator appear
after loop summation in field theory, in other words
it means the substitution of full vertices and propagators
instead of bare ones as it shown in Fig. 2a,b.
\begin{figure}[h]
\begin{center}
\includegraphics*[height=3.5cm]{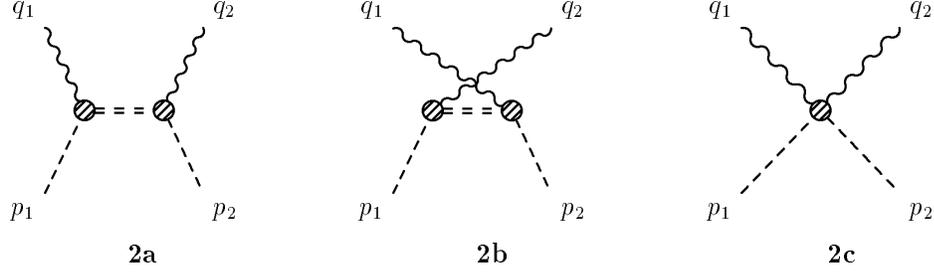}
\end{center}
\caption{Modification of Born terms due to off--shell effects}
\label{MBorn}
\end{figure}
To keep the gauge invariance, one should change the contact term
too (Fig.2c)
but one needs the concrete model for it.
The Ward--Takahashi identity leads to a partial (~but not total~)
cancellation of the loop effects (~see Fig. 3 for illustration
and details below~). Now it's instructive to look into
the full vertex using some field theory.
\begin{figure}[h]
\begin{center}
\includegraphics*[height=2.3cm]{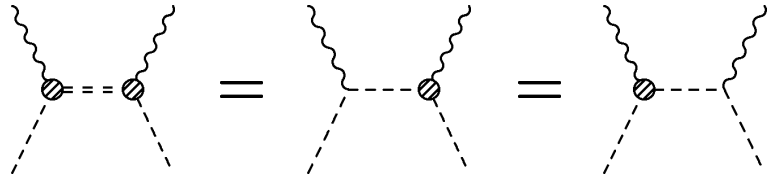}
\end{center}
\caption{Accounting of Ward--Takahashi identity in exchange diagram}
\label{Ward}
\end{figure}

The one--loop contributions in the simplest model
(~interaction of pions with scalar $\sigma$--meson~)
may be seen in Fig. 4. Note that Fig. 3 is equivalent to
contributions of Fig. 4a--c.
Cutting the
$\pi\sigma$ loop in diagram 4a,
we shall have
$\pi\sigma$ real intermediate state with spin $J = 0$ since it is
connected with pion line.
But everyone knows that the partial wave decomposition
of Compton amplitude $\gamma\pi \to\gamma\pi$ starts from  $J = 1$.
The obtained contradiction has simple solution:
contributions from intermediate state  $J = 0$
exist only in single (~gauge non-invariant~) diagrams.
In the total amplitude, the S--wave contributions must
cancel each other, so the off--shell effects should disappear
from gauge--invariant amplitude. We will demonstrate below
by direct calculation
that the S--wave discontinuity on {\bf s} really cancels in sum of
diagrams 4a--d.
\begin{figure}[h]
\begin{center}
\includegraphics*[height=11cm]{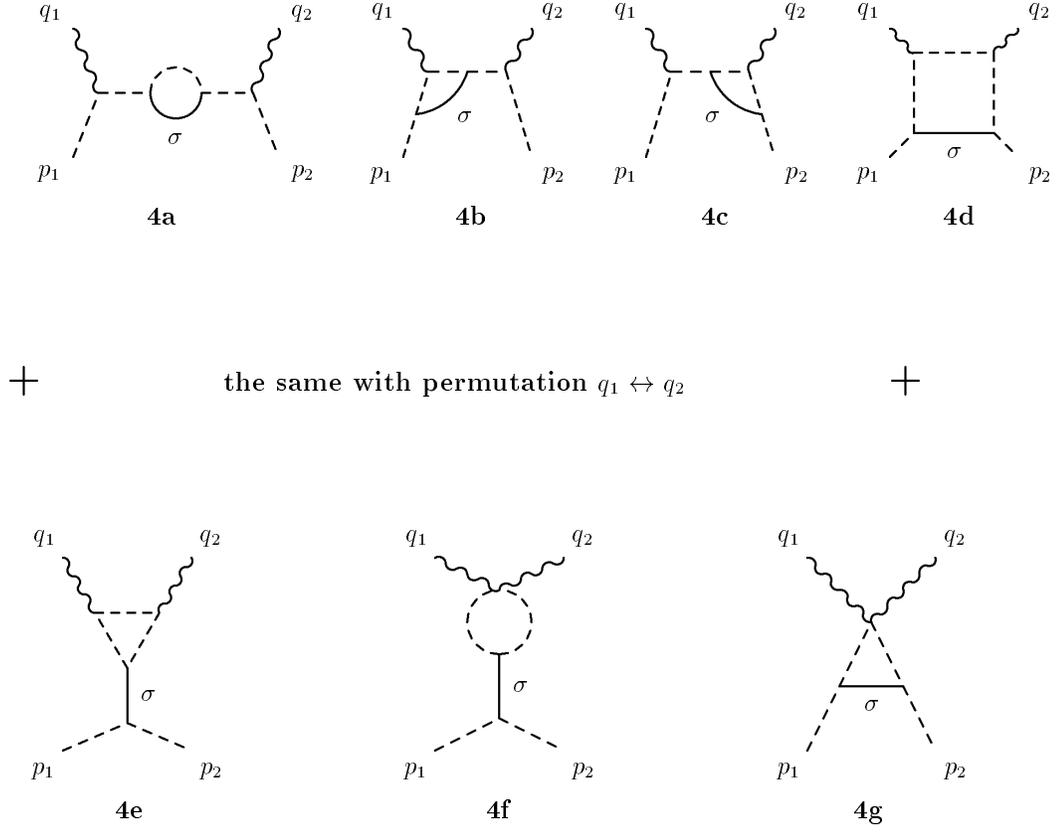}
\end{center}
\caption{One--loop corrections to Born terms in simplest model}
\label{loop}
\end{figure}

So we can argue that the off--shell contributions in Compton
amplitude are absent due to very general reasons not related
with chiral symmetry.

\section{Ward---Takahashi identity}

Let us consider the renormalized irreducible electromagnetic
vertex of charged pion.
\begin{equation}
\Gamma^{\mu}_R = (p_f + p_i)^{\mu}\cdot F(q^2, p^2_f, p^2_i) +
(p_f - p_i)^{\mu}\cdot G(q^2, p^2_f, p^2_i),\ \ \ \
q = p_f - p_i
\label{vert}
\end{equation}
From T--invariance:
\begin{equation}
F(q^2, p^2_f, p^2_i) = F(q^2, p^2_i, p^2_f),\ \ \ \ \
G(q^2, p^2_f, p^2_i) = - G(q^2, p^2_i, p^2_f).
\label{symm}
\end{equation}
The Ward---Takahashi identity \cite{W,T} relates a vertex and
propagator:
\begin{equation}
q_{\mu} \ \Gamma^{\mu}_R =
\Delta^{-1}(p^2_f) - \Delta^{-1}(p^2_i),
\end{equation}
or
\begin{equation}
(p^2_f - p^2_i)\cdot F(q^2, p^2_f, p^2_i)
+ q^2 \cdot G(q^2, p^2_f, p^2_i) =
\Delta^{-1}(p^2_f) - \Delta^{-1}(p^2_i) .
\label{WT1}
\end{equation}
Here
$\Delta(p^2)$ is the renormalized irreducible propagator
of $\pi$-- meson and $\mu = m_{\pi}$.
\begin{eqnarray}
\Delta^{-1}(p^2) &=& p^2 - \mu^2 - J(p^2) =
p^2 - \mu^2 - (p^2 - \mu^2)^2 \cdot \tilde{J}(p^2),
\nonumber\\
   J(\mu^2) &=& J^{\prime}(\mu^2) = 0
\label{prop}
\end{eqnarray}
If the photon and one of pions are on the mass--shell,
the Ward---Takahashi identity reduces to the simple relation.
\begin{equation}
\Delta(p^2_f)\cdot F(0, p^2_f, \mu^2)
= \frac{1}{p^2_f - \mu^2}
\label{WT2}
\end{equation}
Note that an amplitude we are concern
$\gamma\pi^+\to\gamma\pi^+$ for real photons will not
contain the formfactor G, which is accompanied by the
momentum  $q^{\mu}$ and will disappear after multiplying by
polarization vector.

\section{Amplitude with full vertices and propagators}

We consider the Compton scattering by pion:
\begin{center}
$\gamma(\mu,q_1)\ + \pi^+(p_1)\ \  \to \
\gamma(\nu,q_2)\
 + \ \pi^+(p_2)$,\\
\ \ \ \ \ $s = (p_1 + q_1)^2,\ \ \ \ \ t = (q_1 - q_2)^2,\ \ \ \ \
u = (q_1 - p_2)^2$,\ \ \ \ \ \ \ $q_1^2 = q_2^2 = 0$
\end{center}
The Compton amplitude for spinless particle
has a standard decomposition which includes two invariant functions,
free of kinematical singularities and zeros.
\begin{equation}
T^{\mu\nu} = \sum_{i=1}^2 \ \ L^{\mu\nu}_i \cdot B_i(t,s,u)
\end{equation}
It is convenient to use  the following
independent momenta:
$q_1,\ q_2,$ and $ P~=~(p_1 + p_2)/2$.
The gauge--invariant tensor structures are of the form:
\begin{eqnarray}
&& L^{\mu\nu}_1 = q_2^{\mu} \ q_1^{\nu} - (q_1 q_2)\ g^{\mu\nu}
\nonumber\\
&& L^{\mu\nu}_2 = - P^{\mu} P^{\nu} (q_1 q_2) +                    q
 P^{\mu} q_1^{\nu} (q_2  P) +
q_2^{\mu}  P^{\nu} (q_1  P) - g^{\mu\nu} (q_1  P)(q_2  P).
\label{tens}
\end{eqnarray}

Let us start from the standard Born amplitude for point--like
pions.
\begin{equation}
T^{\mu\nu} = e^2\left\{
- (2 p_1 + q_1)^{\mu}\frac{1}{s - \mu^2}(2 p_2 + q_2)^{\nu}
- (2 p_2 - q_1)^{\mu}\frac{1}{u - \mu^2}(2 p_1 -q_2)^{\nu}
+ 2 g^{\mu\nu} \right\}
\end{equation}
One can easily verify the gauge--invariance of this expression.
\begin{equation}
q_{1 \mu} T^{\mu\nu} = q_{2 \nu} T^{\mu\nu} = 0
\label{GI}
\end{equation}

To account the off--shell
effects, we have modify both vertices and propagators,
since the same physical effects generate the full vertices and
propagators.\\
Vertices (see (\ref{vert})):
\begin{equation}
(2 p_1 + q_1)^{\mu} \to (2 p_1 + q_1)^{\mu} F(0,s,\mu^2) -
q_1^{\mu} G(0,s,\mu^2)  \ \  \ \mbox{and so on.}
\end{equation}
Propagator:
\begin{equation}
\frac{1}{s - \mu^2} \to \Delta(s)\ \ \
\end{equation}
One can see that the contact term should be changed too
to keep gauge invariance. Introducing some unknown
generalized contact term  $T_P^{\mu\nu}$ we have the
modified Born contribution (~here the symmetry properties
(\ref{symm}) were taken into account~)
\begin{eqnarray}
\frac{1}{e^2} T^{\mu\nu} =
&-& \left[ (2 P + q_2)^{\mu}\ F(s) + q_1^{\mu}\ G(s) \right] \
\Delta(s) \
\left[ (2 P + q_1)^{\nu}\ F(s) + q_2^{\nu}\ G(s) \right] \
\nonumber \\ &-&
\left[ (2 P - q_2)^{\mu}\ F(u) - q_1^{\mu}\ G(u) \right] \
\Delta(u) \
\left[ (2 P - q_1)^{\nu}\ F(u) - q_2^{\nu}\ G(u) \right] \
\nonumber \\ &+& T_P^{\mu\nu}\\
F(s) &\equiv& F(0,s,\mu^2),\ \ \ \ \  G(s) \equiv G(0,s,\mu^2).
\label{tot}
\end{eqnarray}
Having in mind the Ward---Takahashi identity (\ref{WT2})
we can rewrite the amplitude:
\begin{eqnarray}
\frac{1}{e^2} T^{\mu\nu} =
&-&(2 P + q_2)^{\mu}\ \ \frac{F(s)}{s - \mu^2}\ \ (2 P + q_1)^{\nu}\
- \ (2 P - q_2)^{\mu}\ \ \frac{F(u)}{u - \mu^2} \ \ (2 P - q_1)^{\nu}
\nonumber \\
&+& T_P^{\mu\nu} \  + \
(\mbox{terms} \sim q_1^{\mu} \ \mbox{or} \ q_2^{\nu})
\label{tot2}
\end{eqnarray}

So we can see that the Ward---Takahashi identity
leads to partial cancellation of the off--shell effects
(~see Fig. 3~). As for the generalized contact term  $T_P^{\mu\nu}$,
its form can not be restored unambiguously from the
gauge invariance requirement because we always can add
a gauge--invariant expression to any obtained answer.

\section{One--loop calculations}

Dressing of $\pi$--meson (~i.e. turning of bare propagator into full~)
proceeds due to the three--pion intermediate state. It is
generally believed that the three--pion state is saturated by
the quasi--two--particle $\pi\rho$ and $\pi\sigma$ ones.
To demonstrate the off--shell effects cancellation
\footnote{By the way, a cancellation of the one--loop off--shell
corrections in Compton amplitude was noted a long time ago
\cite{Volkov} in the chiral model. Note that in chiral models there are
definite relations between parameters in contrast to our
calculations.}
we will use the simplest model: interaction of
$\pi$--mesons with a scalar $\sigma$--meson
(nevertheless, it's not the $\sigma$--model ).
\begin{equation}
{\cal L}_{int} = g\ \sigma (x)\vec\pi(x)\vec\pi (x)
\end{equation}

All the one--loop contributions are depicted in Fig. 4.
The diagrams  4e  and 4f have no relation with the off--shell
corrections and are the gauge--invariant together (it's seen due to
presence of the pole $1/(t-m^2)$), so we can forget about them.
We shell consider the discontinuity  of amplitude by s,
originated from the  $\pi\sigma$
intermediate state, only diagrams 4a--4d have such
discontinuities. These contributions are of the form:
\begin{eqnarray}
\Delta_s T^{\mu\nu} = i\ f\ \int d^4 l\
\delta(l^2-\mu^2)&&\delta((l+Q)^2-m^2)\cdot   \nonumber\\
                          \Bigg\{
\frac{(2p_1+q_1)^{\mu}(2p_2+q_2)^{\nu}}{(Q^2-\mu^2)^2}
&-&
\frac{(2l+q_1)^{\mu}(2p_2+q_2)^{\nu}}{(Q^2-\mu^2)[(l+q_1)^2-\mu^2]}\ -
\nonumber\\
-\ \frac{(2p_1+q_1)^{\mu}(2l+q_2)^{\nu}}{(Q^2-\mu^2)[(l+q_2)^2-\mu^2]}
&+&
\frac{(2l+q_1)^{\mu}(2l+q_2)^{\nu}}{[(l+q_1)^2-\mu^2][(l+q_2)^2-\mu^2]}
                           \Bigg\}.
\end{eqnarray}
Here $f=e^2g^2/(2\pi)^2$,\ \ $\mu=m_{\pi}$,\ \ $m=m_{\sigma}$,\ \
$Q = p_1 + q_1$,\ \ $Q^2 = s$.\\
Passing to the set $q_1, q_2, P$\ , let us rewrite
the discontinuity in another form.
\begin{eqnarray}
\Delta_s T^{\mu\nu} &=& i\ f\ \int d^4 l\ \delta(l^2-\mu^2)\
\delta((l+Q)^2-m^2)\cdot \frac{1}{4(lq_1)(lq_2)} \cdot
\nonumber\\
&&                          \bigg\{
4(lq_1)(lq_2)
\frac{(2P+q_2)^{\mu}(2P+q_1)^{\nu}}{(s -\mu^2)^2}\ -\
2(lq_2)
\frac{(2l+q_1)^{\mu}(2P+q_1)^{\nu}}{(s -\mu^2)}\ -\
\nonumber\\
&&\ -\ 2(lq_1)
\frac{(2P+q_2)^{\mu}(2l+q_2)^{\nu}}{(s -\mu^2)}\ + \
(2l+q_1)^{\mu}(2l+q_2)^{\nu}
                           \bigg\}
\label{dis}
\end{eqnarray}
One can verify easily that this expression is gauge--invariant.
The subsequent actions consist in projection  of (\ref{dis})
onto the tensor structure of interest and calculation of
 arising integrals. All this is rather standard so we shall
omit some details. The answer will be expressed in term of
the following scalar integrals.
\begin{eqnarray}
I_0 &=& \int d^4 l\ \delta(l^2-\mu^2)\ \delta((l+Q)^2-m^2)
\nonumber\\
J_1 &=& \int d^4 l\ \delta(l^2-\mu^2)\ \delta((l+Q)^2-m^2)
\frac{1}{(lq_1)} \nonumber\\
J_2 &=& \int d^4 l\ \delta(l^2-\mu^2)\ \delta((l+Q)^2-m^2)
\frac{1}{(lq_2)}  \nonumber\\
V &=& \int d^4 l\ \delta(l^2-\mu^2)\ \delta((l+Q)^2-m^2)
\frac{1}{(lq_1)(lq_2)}
\label{}
\end{eqnarray}

\noindent
\underline{The $q^{\mu}_2 q^{\nu}_1$ term in (\ref{dis})}.

The off--shell corrections of diagrams 4a--4c are very simple and
we will write the answer for them,
not cancelling different terms.
\begin{equation}
\Delta_s T_{abc} = if \bigg\{
\frac{I_0}{(s-\mu^2)^2} - \frac{I_0}{(s-\mu^2)^2}
- \frac{I_0}{(s-\mu^2)^2}   \bigg\},
\label{abc}
\end{equation}
Here three terms in brackets correspond to diagrams
4a---4c. Calculating the integral, we have
\begin{equation}
I_0 =  \pi\ K\ \theta(s-(m+\mu)^2) ,\ \ \ \ \ \ \mbox{where} \
K^2 = \frac{[s-(m+\mu)^2][s-(m-\mu)^2]}{4s}.
\label{}
\end{equation}

 K is a momentum of $\pi\sigma$ state
in the center mass system. Since every term in (\ref{abc})
is proportional to CM momentum they are the S--wave contributions.
We see that first term cancels with second or third,
which corresponds to Ward--Takahashi identity (\ref{WT2}).

Completely the coefficient at $q^{\mu}_2 q^{\nu}_1$
(with accounting that $J_2 = J_1$) is:
\begin{equation}
T_1 = -\frac{if}{4}(\ K_I\cdot I_0 + K_J\cdot J_1 + K_V\cdot V\ ).
\label{coe1}
\end{equation}
The coefficients, calculated with help of REDUCE:
\begin{eqnarray}
K_I &=& \frac{ts}{4(s-\mu^2)^2\Delta^2}
                      \big[
t^3s+t^2(\mu^4 - 6\mu^2 s +5s^2) + 4t(-\mu^6 + 4\mu^4s -5\mu^2s^2
+ 2s^3) + \nonumber \\
&&+ 4(s-\mu^2)^4
                      \big]  \nonumber \\
K_J &=& \frac{t^2(t-4\mu^2)}{8\Delta^2}
(s - \mu^2)(s+\mu^2-m^2)
                               \nonumber \\
K_V &=& \frac{t}{32\Delta^2}
                       \big[
t^3 (-\mu^4+2\mu^2m^2-m^4+2m^2s-s^2) \nonumber \\
&&+ 2 t^2\mu^2 (3\mu^4-4\mu^2m^2-2\mu^2s+2m^4-4m^2s-s^2)
\nonumber \\
&&+ 8t\mu^2 (-\mu^6+3\mu^2s^2-3s^3) -8\mu^2(s-\mu^2)^4
                         \big]   .
\label{}
\end{eqnarray}
Here $\Delta = t(su-\mu^4)/4$ .\\
It is convenient to make the further calculations
in the center mass system $\vec Q = 0$, using s and
scattering angle $c = cos\ \theta$ as the variables.
The remaining scalar integrals are:
\begin{eqnarray}
J_1 &=& \frac{\pi}{(s-\mu^2)}\ \
ln\ \left( \frac{1+\tau}{1-\tau} \right) =
\frac{\pi}{(s-\mu^2)}\ [\ 2\tau + \frac{2}{3}\tau^3 + \ ...\ ]
\nonumber\\
V &=& \frac{4\pi s}{(s-\mu^2)(s+\mu^2-m^2)}\
[\ 2\tau + \frac{2}{3}\tau^3 (c+2) + \ ...\ ],
\nonumber
\label{}
\end{eqnarray}
where
\[
\tau = - \frac{[(s-(m+\mu)^2)(s-(m-\mu)^2)]^{1/2}}{s+\mu^2-m^2}.
\]
We didn't show here rather complicated expression for V,
restricting ourselves by the threshold expansion of corresponding
logarithm.

After substitution into (\ref{coe1}), we can convince
ourselves that the S--wave contributions, proportional
to K , really cancel and decomposition starts from
D--wave,
\begin{equation}
\frac{1}{if}T_1 = \delta^3\cdot \frac{\pi\sqrt{m\mu}}{3}\cdot
\frac{[8\mu^2+14m\mu+7m^2+c(m^2+2m\mu)]}{m^2\mu(m+\mu)^4(m+2\mu)^3}
+ O(\delta^5),
\label{T1}
\end{equation}
where $\delta=\sqrt{s-(m+\mu)^2}$.

\noindent
\underline{The $P^{\mu} P^{\nu}$ term in (\ref{dis}) }.

After similar calculations we found again that the  S--wave
contributions cancel in the sum of diagrams 4a--4c .
The discontinuity in this structure is of the form
\begin{equation}
\frac{1}{if}T_2 = -\delta^3\cdot \frac{4\pi\sqrt{m\mu}}{3}\cdot
\frac{(1-c)}{m\mu(m+\mu)^2(m+2\mu)^4}
+ O(\delta^5).
\label{T2}
\end{equation}

An additional control of calculations consists in construction
of the center of mass helicity amplitudes from  (\ref{T1}), (\ref{T2})
with accounting of (\ref{tens}):
\begin{eqnarray}
T_{0+,0-} &=& - \left[ B_1 \cdot \frac{s}{2} -
B_2 \cdot \frac{s(s-4\mu^2)}{16}  \right],
\nonumber\\
T_{0+,0+} &=& -  B_2 \cdot \frac{su-\mu^4}{4}.
\label{}
\end{eqnarray}
Partial decomposition of helicity amplitudes is well known:
\begin{equation}
T_{0+,0\pm} = \sum_{J=1}^{\infty}(2J+1)\ T_{\pm}^J(s)\
d^J_{1,\pm1}(cos\ \theta) =
\frac{3}{2} T_{\pm}^1(s)\ (1 \pm cos\ \theta) + ...
\label{hel}
\end{equation}

One can verify that angular dependence of the helicity amplitudes
built from (\ref{T1}), (\ref{T2}) indeed correspond to the
lower partial waves in (\ref{hel}).

\section{Conclusion}

We checked by direct calculation that the off-shell effects
which appear in the exchange diagrams cancel in the total amplitude.
It confirms the general arguments indicated in Introduction
about disappearance of the off--shell contributions
in the total gauge--invariant expression for Compton amplitude.
Note that we did not use the chiral symmetry;
the reason for the disappearance is rather kinematical
and model--independent.

Finally, let us note that the corrections to QED
contribution may be investigated in the cross--channel
 $\gamma\gamma\to\pi^+\pi^-$ near the threshold.
Most of experiments performed up to now  studied
the invariant mass region $M_{\pi\pi}\geq 1$ GeV.
As for the near--threshold region $M_{\pi\pi} < 0.5$ GeV,
where the Born term dominates,  only a few
points of MARK II \cite{M2} there exist. Experiments at DA$\Phi$NE
will allow to investigate
this region more accurately
(see, e.g.,\cite{DAP}), so as a result of our consideration we
have a good news for DA$\Phi$NE:\ there are no
off--shell effects in the Compton amplitude.

\section{Acknowledgments}

I would like to thank Thorsten Ohl for his
excellent package {\bf feynMF} \cite{Ohl}.



\begin{thebibliography}{88}
\bibitem{Low} F.T.Low,\ \ Phys. Rev. {\bf 96} (1954)  1428.
\bibitem{Pet} V.A.Petrun'kin,\ \ Phys. Elem. Part. Atom. Nuclei {\bf 12}
 (1981) 692.
\bibitem{MP} D.Morgan and M.R.Pennington,\ \ Z. Phys. {\bf C 48} (1990) 623.
\bibitem{KS} A.E.Kaloshin and V.V.Serebryakov,\ \
   Z. Phys.{\bf C 64} (1994) 689.
\bibitem{CELLO}  H.Behrend et al. (CELLO Coll.),\ \ Z. Phys. {\bf C 56}
  (1992) 381.
\bibitem{Sher} S.Scherer and H.W.Fearing,\ \
Phys. Rev. {\bf C 51} (1995) 359.
\bibitem{W} J.S.Ward,\ \
Phys. Rev. {\bf 78} (1950) 1824.
\bibitem{T} Y.Takahashi,\ \
Nuovo Cimento {\bf 6} (1957) 370.
\bibitem{Volkov} M.K.Volkov and D.Ebert,\ \ Phys. Lett. {\bf B101}
(1981) 252.
\bibitem{M2} J.Boyer et al. (MARK-II Coll.), \ \ Phys. Rev. {\bf D 42}
 (1990) 1350.
\bibitem{DAP} A.Courau and G.Panchery,\ \ in {\bf The DA$\Phi$NE
Physics Handbook},\ ed. by L.Maiani et al.,\
INFN,\ Frascatti\ (1992), p. 353.\\
S.Bellucci,\ \ ibid,\ p. 419.
\bibitem{Ohl} Thorsten Ohl,\ \ Comp. Phys. Comm. {\bf 90} (1995) 340
(see also CTAN).

\end{thebibliography}
\end{document}